# Accurate Time-segmented Loss Model for SiC MOSFETs in Electro-thermal Multi-Rate Simulation

Jialin Zheng, *Student Member, IEEE*, Zhengming Zhao, *Fellow, IEEE,* Han Xu, *Student Member, IEEE*, Weicheng Liu, *Student Member, IEEE*, Yangbin Zeng, *Member,IEEE*

*Abstract*—Compared with silicon (Si) power devices, Silicon carbide (SiC) devices have the advantages of fast switching speed and low on-resistance. However, the effects of non-ideal characteristics of SiC MOSFETs and stray parameters (especially parasitic inductance) on switching losses need to be further evaluated. In this paper, a transient loss model based on SiC MOSFET and SiC Schottky barrier diode (SBD) switching pairs is proposed. The transient process analysis is simplified by time segmentation of the transient process of power switching devices. The electro-thermal simulation calculates the junction temperature and updates the temperature-related parameters with the proposed loss model and the thermal network model. A multi-rate data exchange strategy is proposed to solve the problem of disparity in timescales between circuit simulation and thermal network simulation. The CREE CMF20120D SiC MOSFET device is used for the experimental verification. The experimental results verify the accuracy of the model which provides guidance for the circuit design of SiC MOSFETs. All the parameters of the loss model can be extracted from the datasheet, which is practical in power electronics design.

## I. INTRODUCTION

In recent years, SiC MOSFET devices have received more and more attention to meet the social demand of high efficiency and high power density in power electronics systems [1],[2]. Compared with Si devices of the same capacity, SiC devices have lower on-resistance, smaller junction capacitance, and can withstand higher operating junction temperatures. These advanced features result in lower device losses and higher switching speeds, which can improve the system efficiency and power density of power electronic converters [3]–[5]. SiC devices are increasingly used in high-power, high-voltage, and high-frequency applications, which necessitate higher efficiency and reliability.

Simulation is an effective way to analyze the efficiency and reliability of power electronic systems [6]–[10]. However, SiC devices are generally considered as ideal switches in the simulation of complex power converters, ignoring the delay and distortion of the actual waveform during switching transient. This has led to a series of concerns. ① It is difficult to accurately analyze and simulate the destructive peak voltages and peak currents in transient processes, which threaten the safe operation of the devices. ② It is difficult to accurately calculate the switching loss and propose methods to reduce it. ③ It is difficult to account for the effect of junction temperature on the device parameters. Existing research results indicate that the change in the junction temperature of power devices is an important factor affecting their reliability and lifecycle [11]. Implementing simulations of power electronics converters that characterize SiC MOSFET switching transient loss and junction temperature variations plays a critical role in the assessment of device efficiency and reliability, as well as system-level modeling and engineering analysis. To achieve this goal, the transient loss model should be developed and used in electro-thermal simulation for junction temperature calculation.

The transient power losses can be calculated by two methods. The first method is the experimental fitting method, which obtains several sets of losses under different working conditions by experiment or datasheet, and fits to get the loss value of the required working condition [12]–[14]. Due to the obvious nonlinear characteristics of the SiC MOSFET, the accuracy of the losses obtained by this fitting is not sufficient. The second method is to calculate the losses by simulating the transient waveform of the device. The switching transient process of power switching devices is a complex coupling of multiple mechanisms of physical processes. There are two main types of modeling methods: physical models [15], [16] and behavioral models[17], [18]. The physical model is based on the physical theory of semiconductors, and mathematically expresses the internal mechanisms of the device and fundamentally constructs the physical model of the device. A representative is the McNutt model[15], which is highly accurate, but has large complexity, long simulation time, poor convergence, and is not suitable for engineering analysis. In [16], a PSpice physical model of SiC MOSFETs is developed, and the modeling focus is shifted from the internal mechanism of the device to the external characteristics, which has a shorter simulation time and

This work was supported by the National Natural Science Foundation of China under Grant 51490680 and Grant 51490683. (Corresponding author: Jialin Zheng) better convergence compared with [15], but still cannot meet the demand of system-level simulation.

The behavioral model is designed to simplify the analysis by segmenting the transient process according to the transient switching characteristics and the switching device state. In [17], the switching transient process of a power Si MOSFET is analyzed by using a time-segmented approach, but the model does not give an accurate analysis of the MOSFET turn-on current overshoot caused by the Schottky barrier diode (SBD) (reverse recovery process can be neglected) junction capacitance. In [18], a time segmentation approach is used to propose an analytical loss model of the converter based on SiC MOSFET and SiC SBD, but the model does not consider the effect of the nonlinear characteristics of $C_{gd}$ on the transient process during the rapid change of $v_{ds}$ and does not model the high frequency oscillation at the end of the transient process.

For the temperature characterization of the device, the RC thermal network or the finite element method (FEM) can be used for modeling. In [19], [20], an electrothermal model of power semiconductor devices based on electric-thermal field coupling was constructed in FEM software. Nevertheless, the lack of an accurate electrical model makes it difficult to accurately calculate the power losses of the device, which in turn affects the accuracy of the temperature field solution. In [21]–[24], RC thermal networks were introduced into electrical simulation platforms such as Spice and Saber through the concept of thermal-electrical analogy to construct an electro-thermal model based on electrical-thermal path coupling. This method is computationally more efficient than the FEM analysis method, but Saber and Spice simulations of transient circuits have the problems of slow simulation speed, small simulation circuit size, and poor convergence. In [25], A modeling method is proposed which used the switching pair as a basic unit and proposed a IGBT transient model.

A transient time-segmented loss model is proposed in this paper. The model can reflect the transient waveform and switching loss of the device more accurately and address the difficulty in extraction of physical parameters. This paper is organized as follows. Section II gives the parameters definition and description for the double-pulse test (DPT) circuit to analyze the transient loss of SiC MOSFET and SiC SBD switching pair. Section III proposes the time-segmented loss model and thermal model. Section IV introduces the electro-thermal simulation method based on the proposed loss model. The model proposed is validated in Section V by the operating characteristics of the double-pulse test circuit of Cree SiC MOSFET (CMF20120D) and SiC SBD (C4D30120D). The conclusion is drawn in Section VI.

## II. MODEL DESCRIPTION OF SiC MOSFET AND SiC SBD SWITCHING CELL

In order to model the transients of SiC MOSFET and SiC SBD switching pair, this section uses a double-pulse test (DPT) circuit to analyze and model the transient process of SiC MOSFET in different stages. The parameters and descriptions for the DPT circuit are discussed in details as follows.

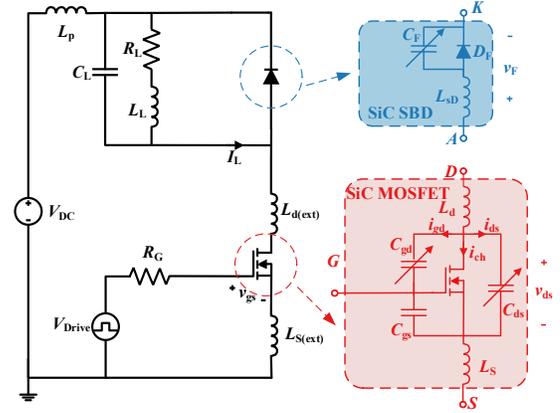

Fig. 1. Equivalent model of the double pulse testing circuit.

### A. Parameters Definition

The DPT equivalent circuit model, which takes into account the stray parameters, is shown in Fig 1. The dc bus voltage is $V_{DC}$, the load inductance $L_L$ is equivalent to the ideal current source $I_L$ and the equivalent parallel capacitance is $C_L$, the output voltage $V_{Drive}$ of the drive circuit is equivalent to an ideal square wave voltage source that jumps between the high-level $V_{CC}$ and the low-level $V_{EE}$. $R_g$ is the resistance of the driving circuit. $L_p$ is the parasitic inductance of the dc bus.

For SiC MOSFET, the equivalent model includes ideal MOSFET, reflecting the static characteristics of SiC MOSFET; three parasitic capacitances, namely gate-source capacitance $C_{gs}$, gate-drain capacitance $C_{gd}$, and drain-source capacitance $C_{ds}$, which affect the transient characteristics of SiC MOSFET. $L_d$ and $L_s$ are drain and source stray inductances. Since the gate stray inductance is generally small, the effect on the transient waveform is not significant and can be ignored.

For SiC SBD, the model includes ideal diode $D_F$, reflecting the static characteristics of SiC SBD; parasitic junction capacitance $C_F$, which affects the transient characteristics of SiC SBD; series stray inductance $L_{sD}$, which can be ignored.

It should be noted that $C_{gd}$, $C_{ds}$, and $C_F$ are non-linear capacitors. The capacitance of the above-mentioned nonlinear capacitors can be linearized piecewise in addition to the $C_{gd}$ during the rapid change of $v_{ds}$ to simplify nonlinearities[26]. Therefore, the capacitance can be divided into different values according to the voltage level, The parameter extraction method of nonlinear capacitance will be introduced in Section V.

Since $L_{sD}$ can be ignored, the equivalent capacitance of the load inductance $C_L$ is considered to be in parallel with the SiC SBD junction capacitance $C_F$. the equivalent junction capacitance of SBD $C_{F(eq)}=C_F+C_L$, the MOSFET input capacitance $C_{iss}=C_{gd}+C_{gs}$, and the output capacitance $C_{oss}=C_{gd}+C_{ds}$. The total stray inductance of the main loop $L_{stray}=L_s+L_d+L_p$.

### B. DPT Circuit Description

The equivalent circuit can be quantitatively described by the following equations.

The KVL in the gate drive circuit:

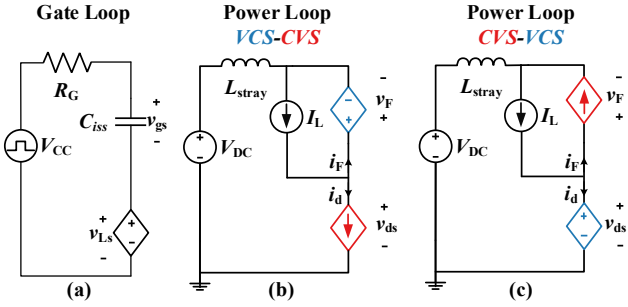

Fig. 2. Equivalent circuits during SiC MOSFET turn-on transition. (a) gate loop. (b) power loop-I. (c) power loop-II.

$$V_{\text{Drive}} = v_{Ciss}(t) + R_G i_g(t) + L_S \frac{di_d(t)}{dt} \quad (1)$$

where $V_{\text{Drive}}$ has two values. When the MOSFET is turning on, its value is $V_{CC}$; When the MOSFET is turning off, its value is $V_{EE}$, $i_g$ is the current in the gate loop, $i_d$ is the MOSFET drain current.

The KCL in the gate drive circuit:

$$i_g(t) = C_{gs}\frac{dv_{gs}(t)}{dt} + C_{gd}\frac{dv_{gd}(t)}{dt}. \quad (2)$$

where $v_{gs}$ is the gate-source voltage of SiC MOSFET, $v_{gd}$ is the gate-drain voltage of SiC MOSFET, $v_{ds}$ is the gate-drain voltage of SiC MOSFET.

The voltage relationship between capacitors:

$$v_{gs}(t) = v_{gd}(t) + v_{ds}(t). \quad (3)$$

The KVL in the DPT circuit:

$$V_{DC} = v_{ds}(t) + L_{stray}\frac{di_d}{dt} - v_F \quad (4)$$

where $v_F$ represents the voltage drop on the SBD,

$$\begin{cases} v_F = v_{F0} & \text{SBD on} \\ i_{Diode} = -C_{F(eq)}\frac{dv_F}{dt} & \text{SBD off.} \end{cases} \quad (5)$$

The KCL in the DPT circuit:

$$i_L(t) = i_F(t) + i_d(t) \quad (6)$$
$$i_d(t) = i_{ch}(t) + i_{gd}(t) + i_{ds}(t) \quad (7)$$

where $i_F$ is the current through SBD, $i_{ch}$ is determined by $v_{gs}$ and $v_{ds}$,

$$i_{ch}(t) = \begin{cases} g_{fs}[v_{gs}(t) - V_{th}] & v_{gs} > V_{th}, v_{ds} > v_{gs} - V_{th} \\ 0 & v_{gs} \leq V_{th} \\ v_{ds}(t)/R_{on} & v_{gs} > V_{th}, v_{ds} \leq v_{gs} - V_{th} \end{cases} \quad (8)$$

where $g_{fs}$ is the linearized transconductance (A/V); $V_{th}$ is the linearized threshold voltage; $R_{ds(on)}$ is the on-state resistance.

The differential equations composed of (1)-(8) are difficult to solve directly. To reduce the calculation, this paper proposed a transient time-segmented model to represent the transient behavior of the switching device in Fig 1. The proposed model is detailed in Section III.

## III. ELECTRO-THERMAL MODEL OF SiC MOSFET AND SiC SBD SWITCHING PAIR

The electro-thermal model is divided into two parts to describe the electrical and thermal behavior of SiC MOSFET and SBD: The electrical model is used to calculate the losses of the MOSFET and SBD. The thermal model is used to calculate the device temperature and to modify the temperature-related parameters of the electrical model according to the temperature.

The total power losses of SiC MOSFET and SBD mainly includes two parts: conduction loss and switching loss. The conduction loss is easy to calculate and will not be described in detail here. This section mainly describes the loss and electrical behavior of MOSFET and SBD during the turn-on and turn-off process. The thermal modeling method and the temperature correction of the parameters are also explained in this section.

The equivalent model of DPT is decoupled to gate loop and power loop shown in Fig 2. The power loop is divided into two modes, CVS-VSC and VCS-CVS, according to the behavior of MOSFETs and SBDs.

### A. Gate Loop Analysis

The time duration of the following stage can be calculated by analyzing the gate loop shown in Fig 2 (a). The voltage of the $L_s$ can be used as a controlled source to decouple the control loop from the power loop.

$$v_{Ls} = L_s \frac{di_d(t)}{dt} \quad (9)$$

Gate loop power supply $V_{drive}$ controls the on/off of the MOSFET by charging and discharging $C_{iss}$ through $R_G$. The voltage relationship in the gate loop can be expressed as

$$V_{\text{Drive}} = R_G i_g + v_{Ciss} + v_{Ls} \quad (10)$$

where $v_{Ciss}$ represents the voltage on the capacitance $C_{iss}$.

The current of the control loop charges and discharges $C_{iss}$. According to (2) and (3), the relationship of the currents can be expressed as

$$i_g = C_{iss}\frac{dv_{gs}(t)}{dt} - C_{gd}\frac{dv_{ds}(t)}{dt}. \quad (11)$$

### B. Turn-on Power Loop Analysis

The SiC MOSFET and SiC SBD switching loss can be calculated as

$$\begin{aligned} E_{\text{loss}}(\text{MOS}) &= \int v_{ds}(t) \cdot i_d(t) dt \\ E_{\text{loss}}(\text{SBD}) &= \int v_F(t) \cdot i_F(t) dt \end{aligned} \quad (12)$$

where $E_{\text{loss}}(\text{MOSFET})$ represents the loss energy produced by SiC MOSFET, $E_{\text{loss}}(\text{SBD})$ represents the loss energy produced by SiC SBD. The analytical expressions $v_{ds}(t)$, $i_d(t)$, $v_F(t)$ and $i_F(t)$ describing the behavior of SBD and MOSFET are related to the turn-on process, which is divided into multiple stages.

A typical SiC MOSFET turn-on transient waveform is shown in Fig 3. The turn-on transient process can be divided into 8 stages. Table I lists the time duration, the current and voltage expression of SiC MOSFET and SiC SBD, and the loss expressions for each stage in the turn-on process. The specific expressions of each stage are introduced in Appendix A.

*Stage 1* ($t_1$-$t_2$): This is $C_{iss}$ charging stage, The output voltage $V_{drive}$ rises from $V_{EE}$ to $V_{CC}$ at $t_1$, and $v_{gs}$ starts to rise. $v_{gs}$ rises to $V_{th}$ at $t_2$. the drive circuit starts to charge the MOSFET input capacitance $C_{iss}$ at $t_1$. Since $L_s$ is small and the rate of change of the current flowing through Ls is small, the $L_s$ voltage $v_{Ls}$ equals zero.

TABLE I
TURN-ON PROCESS

| Stage | Time Duration | Analytical Expression of SBD and MOSFET | Loss Expression of SBD and MOSFET |
|---|---|---|---|
| $t_1$-$t_2$ | $t_2 - t_1 = R_G C_{iss} \ln[\frac{(V_{CC}-V_{EE})}{(V_{CC}-V_{th})}]$ | $\begin{cases} i_d = 0, \ v_{ds} = v_{F(on)} + V_{DC} \\ i_F = I_L, \ v_F = v_{F0} \end{cases}$ | $\begin{cases} E_{loss}(MOS) = 0 \\ E_{loss}(SBD) = I_L v_{F0}(t_2 - t_1) \end{cases}$ |
| $t_2$-$t_3$ | $t_3 - t_2 = \frac{B+\sqrt{(B^2-4AC)}}{2A}$ $A = (V_{CC}-(V_{gs}(t_3)+V_{th})/2)$ $B = R_G C_{iss}(V_{gs}(t_3)-V_{th})+L_{stray}(I_L/2)$ $C = R_G C_{gd} L_{stray} I_L / 2$ | $\begin{cases} i_d(t)=a_{23}(t-t_2)^3+b_{23}(t-t_2)^2+c_{23}(t-t_2)+d_{23} \\ v_{ds}(t)=V_{DC}+v_F-L_{stray}\frac{di_d(t)}{dt} \\ i_F(t)=I_L-i_d(t), \ v_F(t)=v_{F0} \end{cases}$ | $\begin{cases} E_{loss}(MOS) = (V_{DC}+v_F)\int_{t_2}^{t_3} i_d(t)dt - \frac{1}{2}L_{stray} i_d^2(t)\big|_{t_2}^{t_3} \\ E_{loss}(SBD) = v_{F0} I_L(t_3-t_2) - v_{F0}\int_{t_2}^{t_3} i_d(t)dt \end{cases}$ |
| $t_3$-$t_4$ | $t_4 - t_3 = \frac{\begin{bmatrix} R_G C_{iss}(V_{miller}-V_{gs}(t_3))+ \\ L_{stray}(I_L/2) \end{bmatrix}}{(V_{CC}-(V_{miller}+V_{gs}(t_3))/2)}$ | $\begin{cases} i_d(t)=i_d(t)=\frac{I_L}{2(t_4-t_3)}(t-t_3)+\frac{I_L}{2}, \ v_{ds}(t)=V_{ds0} \\ i_F(t)=I_L-i_d(t), \ v_F(t)=v_{F0} \end{cases}$ | $\begin{cases} E_{loss}(MOS) = \frac{3}{4}V_{ds0}I_L(t_4-t_3) \\ E_{loss}(SBD) = \frac{1}{4}(V_{DC}-V_{ds0})I_L(t_4-t_3) \end{cases}$ |
| $t_4$-$t_5$ | $t_5 - t_4 = \frac{\begin{bmatrix} R_G C_{iss}(V_{gs\_peak}-V_{miller})+ \\ L_{stray}(I_{peak}-I_L) \end{bmatrix}}{(V_{CC}-(V_{gs\_peak}+V_{miller})/2)}$ | $\begin{cases} i_d(t)=a_{45}(t-t_4)^3+b_{45}(t-t_4)^2+c_{45}(t-t_4)+d_{45} \\ v_{ds}(t)=V_{ds0}, \ i_F(t)=I_L-i_d(t) \\ v_F(t)=V_{ds0}-V_{DC}+L_{stray}\frac{di_d(t)}{dt} \end{cases}$ | $E_{loss}(MOS) = V_{ds0}C_{F(eq)}(V_{DC}+v_F-V_{ds0})+I_L V_{ds0}(t_5-t_4)$ $E_{loss}(SBD) = (V_{DC}-V_{ds0})C_{F(eq)}(V_{DC}+v_F-V_{ds0})-$ $\frac{1}{2}L_{stray} i_d^2(t)\big|_{t_4}^{t_5} - I_L V_{ds0}(t_5-t_4) + L_{stray} I_L i_d(t)\big|_{t_4}^{t_5}$ |
| $t_5$-$t_6$ | $t_6 - t_5 = \frac{[-R_G C_{gd}(V_{miller}-V_{th}-V_{ds0})]}{(V_{CC}-V_{miller})}$ | $\begin{cases} i_d = I_L + I_{OS}e^{-\alpha_{on}(t-t_5)}\cos(\omega_{on}(t-t_5)) \\ i_F(t) = I_L - i_d(t) \\ v_F(t) = v_{ds}(t) - V_{DC} + L_{stray} di_d(t)/dt \\ v_{ds}(t) = \frac{(V_{t\_end}-V_{t\_start})}{(t_{end}-t_{start})}(t-t_{start})+V_{start}, t_{start} \leq t < t_{end} \end{cases}$ | $E_{loss}(MOS) = I_L V_{ds(av)}(t_8-t_5) + I_{os}\frac{\alpha_{on}}{\alpha_{on}^2+\omega_{on}^2}V_{ds(av)}$ $E_{loss}(SBD) = (V_{DC}-V_{ds(av)})I_{os}\frac{\alpha_{on}}{\alpha_{on}^2+\omega_{on}^2} -$ $\frac{1}{2}L_{stray}(i_d(t)-I_L)^2\big|_{t_5}^{t_8}$ |
| $t_6$-$t_7$ | $t_7 - t_6 = \frac{-R_G C_{gd}(V_{ds(on)}-(V_{miller}-V_{th}))}{(V_{CC}-V_{miller})}$ | | |
| $t_7$-$t_8$ | $t_8 - t_7 \approx 2 R_G C_{iss}$ | | |

The expression of the gate-source voltage $v_{gs}$ in this stage is

$$v_{gs}(t) = V_{CC} + (V_{EE} - V_{CC})\exp\left(-\frac{t-t_1}{R_G C_{iss}}\right). \quad (13)$$

*Stage 2&3 ($t_2$-$t_3$)*: This is the current rising stage. In this stage, $v_{gs}$ starts to rise from $V_{th}$ to $V_{miller}$, MOSFET enters the saturation region. $i_d$ starts to rise from 0 to $I_L$. and $v_{ds}$ starts to fall from $V_{DC}+v_F$. Due to the stray inductance voltage drop $L_{stray}|di_d/dt|$ and the non-linearity of the SBD junction capacitance $C_F$, $v_{ds}$ will remain unchanged when it drops to a certain value $V_{ds0}$. To simplify the analysis, this paper assumes that $v_{ds}$ is $V_{ds0}$ when the current rises to $I_L/2$. Therefore, this stage will be divided into two stages.

*Stage 2 ($t_2$-$t_3$)*: This is the current rising sub-stage I. In this stage, $i_d$ starts to rise from 0 to $I_L/2$, $v_{gs}$ rises from $V_{th}$ to $V_{gs}(t_3)$. $v_{ds}$ drops from $V_{DC}+v_F$ to $V_{ds0}$, $V_{drop}$ is the difference between $V_{DC}+v_F$ and $V_{ds0}$.

*Stage 3 ($t_3$-$t_4$)*: This is the current rising sub-stage II. In this stage, $v_{ds}$ keeps $V_{ds0}$ and $i_d$ starts to linear rise from $I_L/2$ to $I_L$. $v_{gs}$ rises from $V_{gs}(t_3)$ to $V_{miller}$.

*Stage 4 ($t_4$-$t_5$)*: This is the current rising stage III. At $t_4$, the MOSFET current id rises to $I_L$, the SBD current $i_F$ drops to 0, and the SBD turns off. Although the reverse recovery process of SiC SBD is negligible, the $i_d$ overshoots due to the reverse voltage and reverse charging of its junction capacitance. $i_d$ starts to rise from $I_L$ to $I_{peak}$, $v_{ds}$ keeps $V_{ds0}$. $v_{gs}$ rises from $V_{miller}$ to $V_{gs\_peak}$.

*Stage 5 ($t_5$-$t_6$)*: This is the voltage falling stage I. $v_{ds}$ drop from

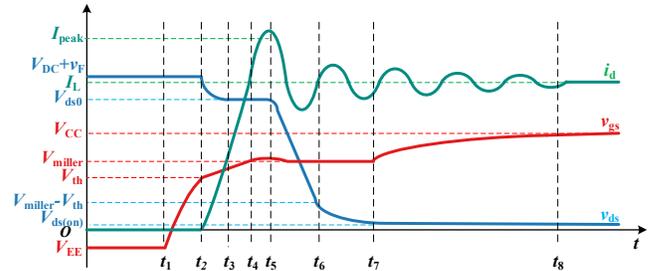
Fig. 3. Turn-on transient waveforms of SiC MOSFET.

$V_{ds0}$ to $V_{miller}-V_{th}$ linearly. $v_{gs}$ is approximately $V_{miller}$ and remains unchanged.

*Stage 6 ($t_6$-$t_7$)*: This is the voltage falling stage II. The gate-drain voltage $v_{ds}$ drop from $V_{miller}-V_{th}$ to $V_{ds(on)}$ linearly, $v_{gs}$ remains approximately $V_{miller}$.

*Stage 7 ($t_7$-$t_8$)*: This is the voltage falling stage III. The gate-drain voltage $v_{ds}$ keeps as $V_{ds(on)}$, The gate-source voltage $v_{gs}$ rises from $V_{miller}$ to $V_{CC}$ exponentially.

*Current Oscillation Stage ($t_5$-$t_8$)*: The drain current of MOSFET $i_d$ enters the oscillation stage from $t_5$ due to the $L_{stray}$ and $C_F$.

In summary, the time durations of each stage can be obtained through (1) and (2), that is, to solve the (10) and (11).

### C. Turn-off Power Loop Analysis

The typical turn-off transient waveform of SiC MOSFET is shown in Fig 4. The turn-off process is the inverse process of the turn-on process. The turn-off transient process can be divided into 5 stages. The time durations, the state expressions of the switching device, and the loss expressions in each stage are shown in Table II. The specific expressions of each stage are introduced in Appendix A.

TABLE II
TURN-OFF PROCESS

| Stage | Time Duration | Analytical Expression of SBD and MOSFET | Loss Expression of SBD and MOSFET |
|---|---|---|---|
| $t_1$-$t_2$ | $t_2 - t_1 = R_G C_{iss} \ln[(V_{CC}-V_{EE})/(V_{CC}-V_{th})]$ | $\{i_d = I_L,\ v_{ds} = V_{ds(on)},\ i_F = 0,\ v_F = -V_{DC} - V_{ds(on)}$ | $\{E_{loss}(\text{MOS}) = I_L V_{ds(on)}(t_2 - t_1),\ E_{loss}(\text{SBD}) = 0$ |
| $t_2$-$t_3$ | $t_3 - t_2 = R_G C_{gd}\dfrac{(V_{miller}-V_{th}-V_{ds(on)})}{(V_{miller}-V_{EE})}$ | $\begin{cases} i_d(t)=I_L, v_{ds}(t)=\dfrac{V_{miller}-V_{th}-V_{ds(on)}}{t_3-t_2}(t-t_2)+V_{ds(on)} \\ i_F(t)=0, v_F(t)=V_{DC}-v_{ds}(t) \end{cases}$ | $\begin{cases} E_{loss}(\text{MOS})=\dfrac{(V_{miller}-V_{th}+V_{ds(on)})}{2}I_L(t_3-t_2) \\ E_{loss}(\text{SBD})=0 \end{cases}$ |
| $t_3$-$t_4$ | $t_4 - t_3 = \dfrac{\begin{bmatrix} R_G C_{iss}(V_{gs}(t_4)-V_{miller})- \\ R_G C_{gd}\Delta V_{ds1}+L_{stray}(I_{t4}-I_L) \end{bmatrix}}{(V_{EE}-(V_{gs}(t_4)+V_{miller})/2)}$ | $\begin{cases} i_F(t)=C_{F(eq)}\dfrac{dv_{ds}(t)}{dt},\ v_{ds}(t)=\dfrac{\Delta V_{ds}}{t_4-t_3}(t-t_3)+V_{miller}-V_{th} \\ i_d(t)=I_L-C_{F(eq)}\dfrac{dv_{ds}(t)}{dt},\ v_F(t)=-V_{DC}+v_{ds}(t)+L_{stray}\dfrac{di_d(t)}{dt} \end{cases}$ | $\begin{cases} E_{loss}(\text{MOS})=\dfrac{(I_L+I_{t4})}{2}\dfrac{(V_{miller}-V_{th}+V_{DC})}{2}(t_4-t_3) \\ E_{loss}(\text{SBD})=\dfrac{-(I_L-I_{t4})}{2}\dfrac{\left(V_{DC}-\dfrac{(V_{miller}-V_{th}+V_{DC})}{2}\right)}{(t_4-t_3)} \end{cases}$ |
| $t_6$-$t_7$ | $t_7 - t_6 = \dfrac{I_{t6}L_{stray}+R_G C_{iss}(V_{miller3}-V_{th})}{(0.5V_{miller}+0.5V_{miller3}-V_{EE})}$ | $\begin{cases} i_d(t)=\dfrac{-I_{t6}}{t_7-t_6}(t-t_6)+I_{t6}, v_{ds}(t)=V_{DC}+V_{OS}\cos(\omega_{off}(t-t_6)) \\ i_F(t)=I_L-i_d(t), v_F(t)=-V_{DC}+v_{ds}(t)+L_{stray}\dfrac{di_d(t)}{dt} \end{cases}$ | $\begin{cases} E_{loss}(\text{MOS})=\dfrac{I_{t6}}{2}\left(V_{DC}+\dfrac{2V_{os}}{\pi}\right)(t_7-t_6) \\ E_{loss}(\text{SBD})=\left(I_L-\dfrac{I_{t6}}{2}\right)v_{F0}(t_7-t_6) \end{cases}$ |
| $t_7$-$t_8$ | $t_8 - t_7 \approx 2 R_G C_{iss}$ | $\begin{cases} i_d(t)=C_{oss}dv_{ds}(t)/dt \\ v_{ds}(t)=V_{DC}+v_{F0}+V_{OS}e^{-\alpha_{off}(t-t_7)}\cos(\omega_{off}(t-t_7)) \\ i_F(t)=I_L-i_d(t), v_F(t)=v_{F0} \end{cases}$ | $\begin{cases} E_{loss}(\text{MOS})=-C_{oss}V_{OS}(V_{DC}+v_{F0})+V_{OS}I_L\dfrac{\alpha_{off}}{\alpha_{off}^2+\omega_{off}^2} \\ E_{loss}(\text{SBD})=I_L v_{F0}(t_6-t_5)+C_{oss}V_{OS}v_{F0} \end{cases}$ |

*Stage 1* ($t_1$-$t_2$): This is $C_{iss}$ discharging stage, $v_{gs}$ drops from $V_{CC}$ to $V_{miller}$. At this stage, the MOSFET is in the linear zone. Neither $v_{ds}$ nor $i_d$ will change. At $t_2$, $v_{gs}$ drops to $V_{miller}$.

*Stage 2* ($t_2$-$t_3$): This is voltage rising stage I. $v_{ds}$ slowly rises from $V_{ds(on)}$ to $V_{miller}-V_{th}$, $v_{gs}$ remains approximately as $V_{miller}$. The SBD is cut-off in this stage, $v_{ds}$ and $v_F$ change slowly, and the SBD junction capacitance current $i_{Diode}$ can be neglected. Therefore, $i_d$ remains approximately as $I_L$ in this stage.

*Stage 3*, ($t_3$-$t_4$): This is voltage rising stage II. the MOSFET enters the saturation region, and $v_{ds}$ rises rapidly. During this stage, $v_F$ drops rapidly, and the SBD junction capacitance discharge current $i_{Diode}$ cannot be ignored, which leads to a slow drop in $i_d$. It is approximately regarded as d $v_F /dt = -dv_{ds}/dt$, and $i_d$ can be expressed as

$$i_d(t)=I_L - C_{F(eq)}\frac{dv_{ds}(t)}{dt}. \quad (14)$$

$v_{ds}$ rises from $V_{miller}-V_{th}$ to $V_{DC}$, $\Delta V_{ds1}=V_{DC}-(V_{miller}-V_{th})$. $i_d$ drops from $I_L$ to $I_{t4}$, $v_{gs}$ decreases from $V_{miller}$ to $V_{gs}(t_4)$ due to $i_d$ falling.

*Stage 4* ($t_4$-$t_5$): $v_{ds}$ rises to $V_{DC}+V_{F0}$ at $t_6$. Due to the rapid decline of $i_d$ through inductance $L_{stray}$, $v_{ds}$ is overmodulated to the peak value $V_{peak}$. $i_d$ drops from $I_{t6}$ to 0, $v_{gs}$ decreases from $V_{gs}(t_6)$ to $V_{th}$.

*Stage 5* ($t_5$-$t_6$): $v_{gs}$ drops to $V_{th}$ at $t_6$ and continue to decline exponentially to $V_{EE}$. The MOSFET enters the cut-off region. It is considered that the SBD starts to conduct at this moment. $v_{ds}$ and $i_d$ come into oscillation stage.

### D. Thermal Model and Parameters Correction

The thermal network is the key module to convert the power loss of the SiC MOSFET into the junction temperature of the device. Fig 5 shows the *1-D* equivalent thermal network model of the SiC MOSFET and SBD. The SiC MOSFET conducts heat conduction from the PN junction to the external environment $T_{amb}$. The power loss passes through the thermal network from the case to the external environment to obtain the case temperature $T_c$, and then passes through the PN junction to the thermal network of the substrate to obtain the junction temperature $T_j$. The temperature $T_j$ of MOSFET and SBD can be expressed as

$$T_j = P_{loss}\sum_{i=1}^{4}\left(\frac{R_{th}(i)}{R_{th}(i)C_{th}(i)+1}\right)+T_C \quad (15)$$

$$T_C = P_{loss}\left(Z_{th}^G + Z_{th}^H\right)+T_{amb}.$$

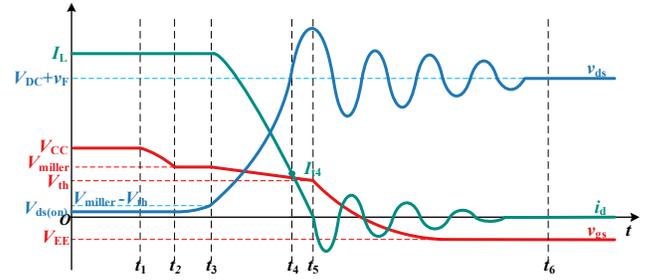

Fig. 4. Turn-off transient waveforms of SiC MOSFET.

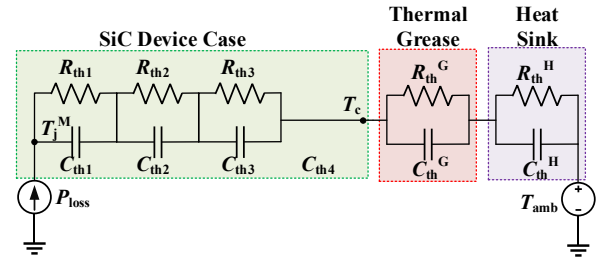

Fig. 5. Thermal calculation circuit of the power electronics device.

The change of junction temperature $T_j$ has a great influence on the value of electrical parameters. The influence of $T_j$ on the characteristics of SiC MOSFET and SiC SBD is mainly manifested in the influence on threshold voltage $V_{th}$, transconductance coefficient $k$, and on-resistance $R_{ds(on)}$. The relationship can be expressed as

$$V_{th} = V_{th0} + a(T_j - T_0) \quad (16)$$

$$k_{fs} = k_0 + b(T_j - T_0) \quad (17)$$

$$R_{ds(on)} = R_{ds(on)0}\left(cT_j^2 + dT_j + e\right) \quad (18)$$

where a-e are coefficients and can be obtained by the curve fitting method.

It is noted that the linear transconductance $g_{fs}$ and threshold voltage $V_{th}$ can be derived as

$$g_{fs} = \frac{2(\lambda^2 + 3\lambda + 3)}{3\lambda(1+\lambda)}\sqrt{k_{fs}I_L}, \lambda = \sqrt{6} \quad (19)$$

$$V_{th} = \sqrt{I_L / k_{fs}} / (1 + k_{fs}) + V_{th0}$$

In addition, since the temperature has a small influence on the capacitance between electrodes, the correlation between the capacitance and temperature is not considered here.

## IV. ELECTRO-THERMAL SIMULATION BASED ON TIME-SEGMENTED LOSS MODEL

### A. Electro-thermal simulation Strategy

The relationship between the electrical model and thermal model of transient switching pair (TSP) can be expressed as follow, as shown in Fig 6. The power losses generated by MOSFET lead to an increase in module temperature, and the temperature changes affect the electrical characteristics of the MOSFET. Therefore, loss and temperature parameters are an important bridge between electrical and thermal simulations. The two models are coupled electrically and thermally by the mutual transfer of temperature and loss. The large difference in the timescales of the two models is a challenge for the interaction of loss and temperature data.

### B. Data exchange between loss model and thermal model

An adaptive data exchange strategy is proposed to solve the problem of disparity in timescales between circuit simulation (ns) and thermal network simulation (ms). The adaptive step $dt_{th}(i)$ is adjusted according to the temperature change $dT(i)$ calculated in the previous step.

$$dt_{th}(i) = t_{th}(i) - t_{th}(i-1) \quad (20)$$

$$dT_{th}(i) = T_{th}(i) - T_{th}(i-1) \quad (21)$$

Temperature exceeding the upper threshold is defined as an event. When the temperature rises, if the temperature rise $dT(i)$ does not exceed the upper threshold, $dt_{th}(i)$ continuously increases by a constant $\xi$. If $dT(i)$ exceeds the upper threshold, i.e., the thermal-level event occurs, $dt_{th}(i)$ maintains the previous step size.

$$\begin{cases} dT_{th}(i) - dT_{th}(i-1) \leq \Delta T \Rightarrow dt_{th}(i) = dt_{th}(i-1) + \zeta \\ dT_{th}(i) - dT_{th}(i-1) > \Delta T \Rightarrow dt_{th}(i) = dt_{th}(i-1) \end{cases} \quad (22)$$

The temperature drop to the lower threshold is defined as another event. As the temperature drops, $dt_{th}(i)$ first changes back to the minimum step size. If $dT(i)$ does not exceed the lower threshold, $dt_{th}(i)$ continuously increases by a constant $\xi$. If the $dT(i)$ exceeds the lower threshold, the thermal-level event occurs and $dt_{th}(i)$ remains the same as the previous step.

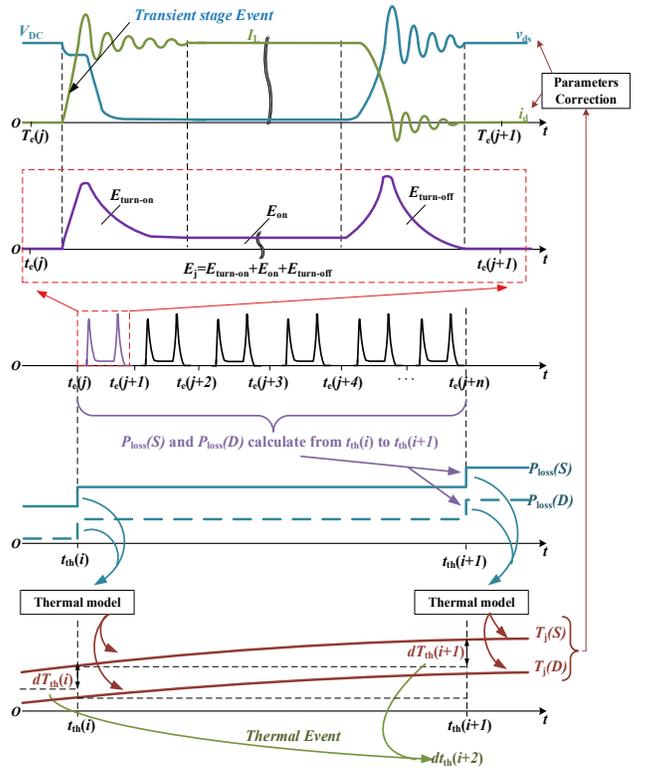

Fig. 6. Proposed electro-thermal simulation overview.

$$\begin{cases} dT_{th}(i) < 0, dT_{th}(i-1) < 0 \Rightarrow dt_{th}(i) = dt_{th}(0) \\ |dT_{th}(i)| - |dT_{th}(i-1)| \leq \Delta T \Rightarrow dt_{th}(i) = dt_{th}(i-1) + \zeta \\ |dT_{th}(i)| - |dT_{th}(i-1)| > \Delta T \Rightarrow dt_{th}(i) = dt_{th}(i-1) \end{cases} \quad (23)$$

With the same simulation accuracy, the adaptive time step increases the simulation efficiency by adjusting the simulation step according to the rate of temperature change, although it increases the complexity compared to a fixed step. The cost of this adaptive step management is that the power loss must be averaged over the circuit simulation time, as shown in Fig 6, due to the large difference in time scales between the device level simulation and the thermal simulation. As such, very small thermal transients (e.g., temperature fluctuations due to a single switching event) will be ignored.

It should be noted that the step size $dt_{th}(i)$ affects the temperature variation $dT(i)$, which has an impact on the device the electrical behavior. Therefore, a reasonable temperature variation threshold requires to be set, and the temperature variation threshold $\Delta T$ set in this paper is 1°C.

## V. EXPERIMENT VERIFICATION

### A. Experiment Parameters Extraction

The DPT platform is shown in Fig 7. The SiC MOSFET (CMF20120D) and SiC SBD (C4D30120D) from Wolfspeed are used as examples for the following experimental verification[27], [28]. The method of extracting the transient model parameters of SiC MOSFET and SiC SBD based on the datasheet is given as follows and their values are shown in Table III. The parameters to be extracted in the SiC MOSFET model include threshold voltage $V_{th}$, transconductance $g_{fs}$, parasitic capacitance $C_{gs}$, $C_{gd}$, and $C_{ds}$.

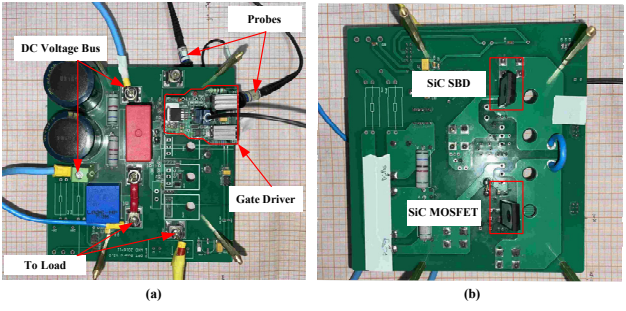

Fig. 7. The double pulse test circuit. (a) Top view. (b) Bottom view.

TABLE III
PARAMETERS OF SiC MOSFET MODEL

| Parameter | Value | Parameter | Value |
|---|---|---|---|
| $C_L$/pF | 26 | $L_s$/nH | 6 |
| $V_{CC}$/V | 20 | $L_d$/nH | 150 |
| $V_{EE}$/V | -5 | $g_{fs}$/S | 4.9 |
| $V_{miller}$/V | 9.66 | $C_{gs}$/nF | 2 |
| $C_{ds0}$/nF | 1.4 | $C_{gd0}$/pF | 571 |
| $C_{ds1}$/pF | 139 | $C_{gd1}$/pF | 15 |
| $C_{ds2}$/pF | 95 | $C_{gd2}$/pF | 11 |
| $V_{th}$ | 5.9 | $R_{G(int)}$/Ω | 5 |
| $C_{F0}$/nF | 1.2 | $v_{F0}$/V | 1.3 |
| $C_{F1}$/pF | 86 | $R_{DS(on)}$/mΩ | 80 |
| $C_{F2}$/nF | 61 | $L_{sD}$/nH | 6.5 |
| $R_{th1}^M$(K/W) | 0.078 | $C_{th1}^M$(Ws/K) | 0.005 |
| $R_{th2}^M$(K/W) | 0.197 | $C_{th2}^M$(Ws/K) | 0.018 |
| $R_{th3}^M$(K/W) | 0.162 | $C_{th3}^M$(Ws/K) | 0.249 |
| $R_{th1}^D$(K/W) | 0.045 | $C_{th1}^D$(Ws/K) | 0.004 |
| $R_{th2}^D$(K/W) | 0.179 | $C_{th2}^D$(Ws/K) | 0.014 |
| $R_{th3}^D$(K/W) | 0.144 | $C_{th3}^D$(Ws/K) | 0.232 |

The threshold voltage $V_{th}$ and transconductance $g_{fs}$ can be obtained from the transfer characteristic curves of SiC MOSFET. The channel current $i_{ch}$ and the gate-source voltage $v_{gs}$ satisfies the relationship

$$i_{ch} = k_{fs}\left(v_{gs} - v_{th0}\right)^2. \qquad (24)$$

where $k_{fs}$ and $v_{th0}$ can be obtained by fitting the transfer characteristic curve of the datasheet.

The MOSFET transfer characteristic curve is linearized at $i_{ch}=I_L/2$ and replaced by a tangent line to simplify the analysis. The linearization relationship between channel current $i_{ch}$ and the gate-source voltage $v_{gs}$ is shown in (8). The threshold voltage $V_{th}$ and transconductance $g_{fs}$ can be calculated from $k_{fs}$ and $v_{th0}$ using (19).

The gate source capacitance $C_{gs}$ is the gate oxide layer capacitance, and its capacitance can be approximated as a constant due to the constant oxide layer thickness; the segmented capacitance values reflect the nonlinearity of the parasitic capacitance $C_{ds}$ and $C_{gd}$ to facilitate the calculation. The extraction method of the SBD junction capacitance $C_{jD}$ is similar to that of $C_{ds}$ and $C_{gd}$.

### B. Loss Model Verification

(1) Switching waveform verification-By varying the magnitudes of the bus voltage and load inductance, the switching transient waveforms are obtained under different voltages and currents for theoretical calculations and experimental measurements, as shown in Fig 8.

In Fig 8, it can be seen that the model proposed in this paper agrees well with the experimental measurements at different voltages and currents. It can be seen that the model proposed in this paper matches well with experimental measurements at different voltages and currents, both for the turn-on process and the turn-off process and can reflect the SiC MOSFET switching transient waveform characteristics in a more detailed way.

(2) Switching loss verification-The accuracy of the loss calculation of the proposed model is verified by changing the external resistance $R_{G(ext)}$ of the drive circuit and the external gate-drain capacitance $C_{gd(ext)}$, respectively. The theoretical calculation results are compared with the experimentally measured results for $V_{DC}$=400V, $I_L$=15A, and the results are shown in Fig 9.

Fig. 9 (a) and (b) show the theoretical calculations and experimental measurements of the turn-off and turn-on losses for $R_G$=10, (c) and (d) show the comparison of the experimentally measured losses and the theoretically calculated losses for different $R_G$ and different $C_{gd}$ conditions.

From Fig. 9 (a) and (b), it can be seen that the turn-on loss is mainly generated in the current-rising stage and voltage-falling stage; the turn-off loss is mainly generated in the current-falling stage and voltage-rising stage. Although the theoretical waveform of this model does not match exactly with the experimental test in the oscillation phase, the oscillation phase has little effect on the turn-on loss and turn-off loss. Under different $R_G$ and $C_{gd}$ conditions, the total loss errors of the theoretical calculation and experiment are less than 6%, and the error comparison is shown in Table IV.

(3) parameters correction verification-Fig 10 (a) shows the

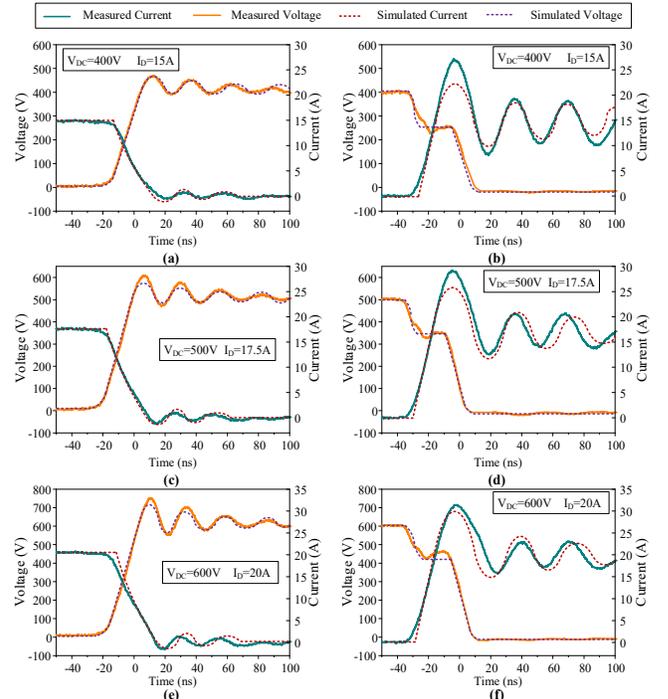

Fig. 8. Analytical calculation and experimental waveforms under different conditions.

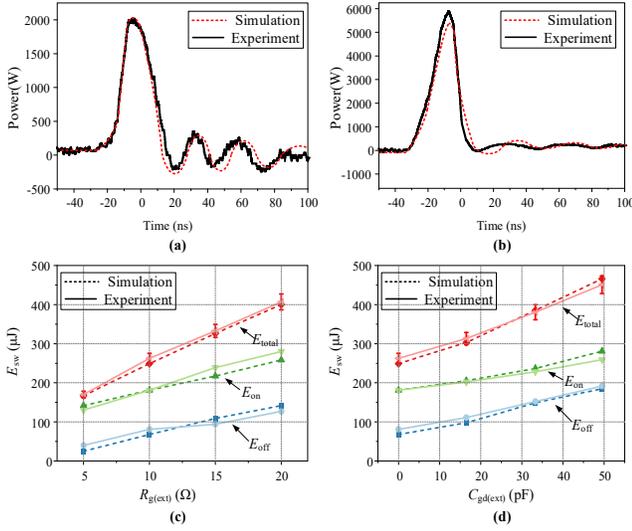

Fig. 9. Analytical calculation and experimental switching losses. (a) turn off loss waveform. (b) turn on loss waveform. (c) with different $R_{g(ext)}$. (d) with different $C_{gd(ext)}$

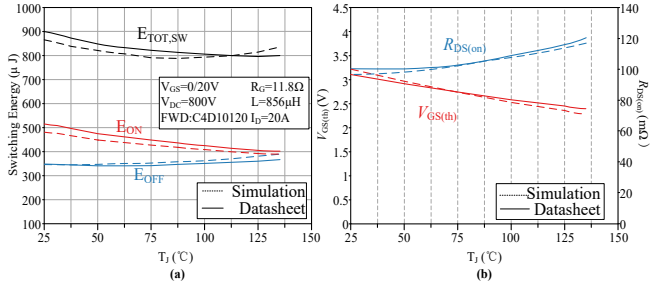

Fig. 10. (a) Simulate and datasheet switching energy versus junction temperature. (b) Simulate and datasheet parameters versus junction temperature.

simulation results of the switching energy of the SiC MOSFETs, which is obtained by simulating the switching energy of SiC MOSFETs at different temperatures to find the power loss of the device. Comparing the simulation results with the datasheet, it is found that the simulation data matches well with the datasheet at lower temperatures (about less than 100 °C), while the error increases at higher temperatures (more than 100 °C), and the high-temperature characteristics of the model need to be improved. The simulation results also show that the effect of junction temperature on the switching loss is not significant, especially the turn-off switching loss does not change with the change of junction temperature. Fig 10 (b) shows the variation of temperature-related parameters of SiC MOSFET, which is consistent with the actual situation.

TABLE IV
COMPARISON OF EXPERIMENTAL LOSS AND SIMULATION LOSS

| | $R_G/\Omega$ | 5 | 10 | 15 | 20 |
|---|---|---|---|---|---|
| Trun-on Loss | Experiment (μJ) | 130.153 | 181.203 | 238.869 | 279.968 |
| | Simulation (μJ) | 141.779 | 181.195 | 217.300 | 258.384 |
| | Error (%) | 8.93% | 0.00% | 9.03% | 7.71% |
| Trun-off Loss | Experiment (μJ) | 40.078 | 81.162 | 94.023 | 126.792 |
| | Simulation (μJ) | 25.117 | 67.869 | 108.952 | 141.745 |
| | Error (%) | 37.33% | 16.38% | 15.88% | 11.79% |
| Total Loss | Experiment (μJ) | 170.231 | 262.365 | 332.892 | 406.76 |
| | Simulation (μJ) | 166.896 | 249.064 | 326.252 | 400.129 |
| | Error (%) | 1.96% | 5.07% | 1.99% | 1.63% |
| | $C_{gd(ext)}/pF$ | 0 | 16.5 | 33.3 | 49.5 |
| Trun-on Loss | Experiment (μJ) | 181.195 | 202.066 | 228.177 | 259.273 |
| | Simulation (μJ) | 181.203 | 205.414 | 236.526 | 280.94 |
| | Error (%) | 0.00% | 1.66% | 3.66% | 8.36% |
| Trun-off Loss | Experiment (μJ) | 81.162 | 111.081 | 152.207 | 191.637 |
| | Simulation (μJ) | 67.869 | 97.778 | 148.859 | 184.97 |
| | Error (%) | 16.38% | 11.98% | 2.20% | 3.48% |
| Total Loss | Experiment (μJ) | 262.365 | 313.147 | 380.384 | 450.91 |
| | Simulation (μJ) | 249.064 | 303.192 | 385.385 | 465.91 |
| | Error (%) | 5.07% | 3.18% | 1.13% | 3.33% |

TABLE V
TEMPERATURE MEASURED TEST CONDITION

| Parameter | Value | Parameter | Value |
|---|---|---|---|
| $V_{DC}$/V | 400 | $R_L/\Omega$ | 5 |
| $L_L$/μH | 1000 | D(0s-2s) | 0.5 |
| D(2s-4s) | 0.4 | D(4s-6s) | 0.3 |
| D(6s-8s) | 0.4 | D(8s-10s) | 0.5 |
| MOSFET | CMF20120D | SBD | C4D30120D |

### C. Thermal Model Verification

(1) Junction temperature verification-In this paper, FLUKE's Ti480 PRO thermal imager is used to measure the device temperature. The ambient temperature is 25 degrees Celsius. To demonstrate the accuracy of the temperature calculation, this paper creates a variety of test conditions by varying the duty cycle of the control signal. To ensure the accuracy of the temperature calculation measurements, the experiment was repeated five times in the same place using a thermal imager and averaged for smaller errors. The thermal image of SiC MOSFET and SiC SBD at 1s are shown in Fig 11, and the temperature measurement and thermal-level simulation are shown in Fig 12 (a).

The maximum error of the SiC MOSFET temperature is 3.2 degrees, and the relative error is 6.17%; the maximum error of the SBD temperature is 2.7 degrees, and the relative error is 6.87%. Modeling of thermal impedance is not the focus of this paper, further investigation is required.

(2) Data exchange step verification-Different data exchange steps can have an impact on the results of temperature calculation and parameter correction. Fig 12 (b) and (c) shows the temperature and step size using two different sizes of fixed steps and the variable step size proposed in this paper. If very small step sizes are used, e.g., 1e-5 s, the dynamics of the temperature change process can be simulated very finely. Conversely, a large number of computational resources are required for calculating the loss power and temperature corrections. If a very large step size is used, e.g., 1e-2 s, the computational effort to update the loss power and temperature corrections can be greatly reduced. However, the calculated temperature differs significantly from that using a small step size (1e-5 s).

The adaptive step size proposed in this paper can compromise the accuracy of temperature calculation and the consumption of computational resources. As can be seen in Fig 12 (b), although the step size is much larger than the fixed step size (1e-5 s), the accuracy is already very close, with a maximum error of 2 degrees Celsius. From Fig 12 (c), it can be seen that for a simulation of 0.1s, if a fixed step size (1e-5 s) is used, the data needs to be updated 1000 times, while the adaptive step size only needs to be updated 26 times. Thermal simulations are often at the minute level or even longer, this adaptive data exchange method can greatly save computational

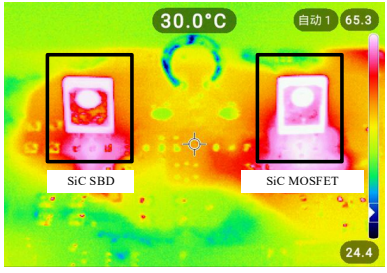

Fig. 11. Thermal image of SiC MOSFET and SiC SBD

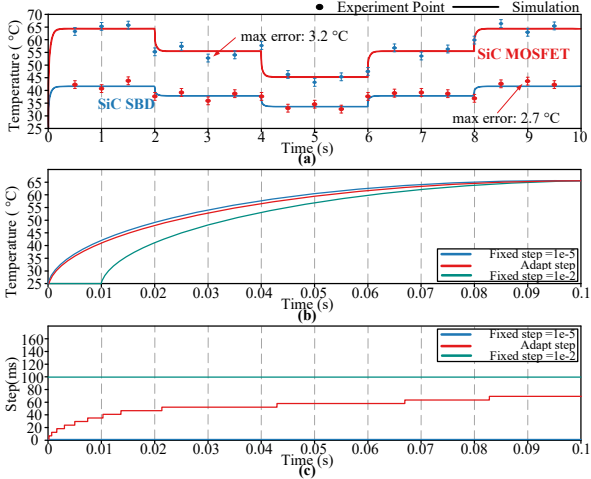

Fig. 12. Analytical calculation and experimental device temperature. Comparison of MOSFET case temperature using different data exchange step. (a) Temperature. (b) Step size.

resources while ensuring computational accuracy subsequently.

## VI. CONCLUSION

In this paper, a transient time-segmented loss model for SiC power switching devices is proposed, which can be used for the simulation of the switching transients, power losses and device junction temperatures of SiC MOSFETs.

1. In this paper, a transient time-segmented loss model based on SiC MOSFET and SiC SBD switching pair is established. The model has considered the impact of displacement current which cannot be neglected when the drain current is low. The model also considers the impact of nonlinear parameters, such as the junction capacitances, the temperature-related parameters, and the oscillation phenomenon caused by stray parameters.

2. The proposed losses model can also calculate the transient waveform and characterize the switching transients of SiC MOSFETs in more details compared with other behavior model of SiC MOSFET. The model parameters can be extracted from the datasheet, which has strong practicality. This model requires only external conditions: bus voltage and load current, and can be applied in all kinds of system-level simulation method with generality.

3. The proposed loss model can also be used in junction temperature calculation. The loss model and thermal network model are used in system-level simulation to compose an electro-thermal simulation platform. And the strategy is proposed where the adaptive data interaction step can greatly save the computational resources while ensuring computational accuracy.

## APPENDIX
### DERIVATIONS OF SWITCHING PROCESS

*A. Turn-on Analysis*

Stage 2 ($t_2$-$t_3$): $v_{gs}$ increases to $V_{gs}(t_3)$ at $t_3$, which is expressed as

$$V_{gs}(t_3) = \frac{I_L}{2g_{fs}} + V_{th}. \quad (25)$$

Meanwhile, $v_{ds}$ decreases to $V_{ds0}$ and the difference between $V_{ds0}$ and $V_{ds}$ is $V_{drop}$. $V_{ds0}$ and $V_{drop}$ are expressed as

$$V_{ds0} = V_{DC} + v_F - V_{drop} = V_{DC} + v_F - L_{stray}\frac{I_L}{2(t_3 - t_2)}. \quad (26)$$

Stage 3 ($t_3$-$t_4$): $v_{gs}$ increases to $V_{miller}$ at $t_4$ when $i_d = I_L$, $V_{miller}$ is

$$V_{miller} = \frac{I_L}{g_{fs}} + V_{th}. \quad (27)$$

Stage 4 ($t_4$-$t_5$): Due to SBD capacitance discharging at this stage, $i_d$ increases from $I_L$ and reaches to $I_{peak}$ at $t_5$. $I_{peak}$ is

$$I_{peak} = I_L + \frac{2C_{F(eq)}(V_{DC} + v_F - V_{ds0})}{t_5 - t_4}. \quad (28)$$

According to (8), $V_{gs\_peak}$ is

$$V_{gs\_peak} = \frac{I_{peak}}{g_{fs}} + V_{th}. \quad (29)$$

Stage 6 ($t_6$-$t_7$): the SiC MOSFET is in the saturation region at this stage. $V_{ds(on)}$ is

$$V_{ds(on)} = I_L R_{ds(on)}. \quad (30)$$

*B. Turn-off Analysis*

Stage 3 ($t_3$-$t_4$): $i_d$ increases to $I_{t4}$ at $t_4$. $I_{t4}$ is expressed using (14) as

$$I_{t4} = I_L - C_{F(eq)}\frac{\Delta V_{ds1}}{t_4 - t_3}. \quad (31)$$

According to (8), $V_{gs}(t_6)$ is

$$V_{gs}(t_4) = \frac{1}{g_{fs}}\left(I_{t6} - C_{oss}\frac{\Delta V_{ds3}}{t_6 - t_5}\right) + V_{th}. \quad (32)$$

Stage 4 ($t_4$-$t_5$): $v_{ds}$ increases to $V_{peak}$ at $t_7$. $V_{peak}$ is expressed as

$$V_{peak} = V_{DC} + v_{F0} + L_{stray}I_{t6}/(t_7 - t_6). \quad (33)$$

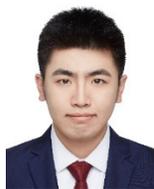

**Jialin Zheng.** (Student Member, IEEE) received the B.S. degree in electrical engineering in 2019 from Beijing Jiaotong University, Beijing, China. Since 2019, he has been working toward the Ph.D. degree in electrical engineering at the Department of Electrical Engineering, Tsinghua University, Beijing, China. His research interests include real-time simulation, in power electronics.

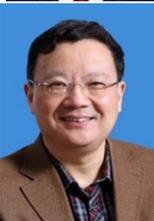

**Zhengming Zhao** (Fellow, IEEE) received the B.S. and M.S. degrees in electrical engineering from Hunan University, Changsha, China, in 1982 and 1985, respectively, and the Ph.D. degree in electrical engineering from Tsinghua University, Beijing, China, in 1991. He is currently a Professor with the Department of Electrical Engineering, Tsinghua University. His research interests include high-power conversion, power electronics and motor control.

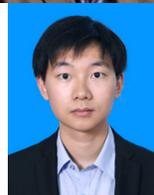

**Han Xu** (Student member, IEEE) received the B.S. degree in electrical engineering in 2021 from Tsinghua University, Beijing, China. Since 2021, he has been working toward the master degree in electrical engineering at the Department of Electrical Engineering, Tsinghua University, Beijing, China. His research interests include simulation of power electronic systems.

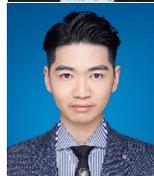

**Weicheng Liu** (Student member, IEEE) received the B. E. degree in electrical engineering from Southeast University, China, in 2018, and M.S degree in electrical engineering from Southeast University in 2021. He is currently working towards the Ph.D. degree in electrical engineering from Tsinghua University, China. His current research interests include power system simulation methodology and technique.

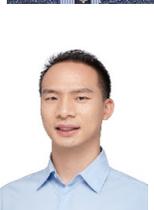

**Yangbin Zeng** (Member, IEEE) received the B.Sc. degree in Building Electrical and Intelligent from Xiangtan University, Xiangtan, China, in 2015, the Ph.D. degree in electrical engineering from Beijing Jiaotong University, Beijing, China, in 2021. Now, he is a postdoctoral researcher in Department of Electrical Engineering, Tsinghua University, Beijing. His current research interests include real-time simulation.